\documentclass[aps,prd,superscriptaddress,twoside,twocolumn,nofootinbib,%
10pt,showpacs,floatfix]{revtex4-2}

\usepackage{color}
\usepackage{amsmath,amssymb}
\usepackage{graphicx,bm}
\usepackage{ulem}
\usepackage{slashed}
\usepackage{dcolumn}
\usepackage{epsfig}
\usepackage{multirow}

\allowdisplaybreaks

\begin{document}


\title{Mixing mechanism for the $J^{P}=0^{+}$ mesons}


\author{Hungchong Kim}%
\email{bkhc5264@korea.ac.kr}
\affiliation{Center for Extreme Nuclear Matters, Korea University, Seoul 02841, Korea}

\author{K. S. Kim}%
\affiliation{School of Liberal Arts and Science, Korea Aerospace University, Goyang, 412-791, Korea}

\date{\today}


\begin{abstract}
There are three scalar nonets in the Particle Data Group (PDG), one of which includes [$a_0(980), K_0^*(700)$],
another includes [$a_0(1450), K_0^*(1430)$], and the third includes [$a_0(1710), K_0^*(1950)$].
Motivated by Ref.~\cite{Black:1999yz}, we examine an alternative mixing mechanism that could potentially
explain the small mass difference between the $a_0 (1450)$ and $K_0^* (1430)$.
According to the tetraquark mixing model, two types, distinguished by their color-spin structures,
are necessary to describe the tetraquark structure of the two nonets containing [$a_0(980), K_0^*(700)$] and [$a_0(1450), K_0^*(1430)$].
Considering the color-spin structures, we argue that the mixing mechanism generating $a_0(1450)$ and $K_0^* (1430)$ on the one hand,
and $a_0(1710)$ and $K_0^* (1950)$ on the other hand might be relevant for resolving the small mass difference.
We also discuss the limitations of other mixing mechanisms that generate the two nonets involving [$a_0(980),K_0^*(700)$] and [$a_0(1450)$, $K_0^* (1430)$] or
[$a_0(980),K_0^*(700)$] and [$a_0(1710)$, $K_0^* (1950)$].
\end{abstract}
\maketitle

\section{Introduction}
\label{introduction}

It is well known that the SU$_f$(3) quark model~\cite{Gell-Mann:1964ewy},
which utilizes the three light quarks, $q = u, d, s$, as a fundamental representation, can classify the lowest-lying hadrons.
The pseudoscalar mesons ($J^P=0^-: \pi, K, \eta, \eta^\prime$) and the vector mesons,
($J^P=1^-: \rho, K^*, \omega, \phi$), which separately form a flavor nonet, $\bm{8}_f\oplus\bm{1}_f=\bm{9}_f$,
can be understood very well from the $q\bar{q}$ structure.
For mesons containing heavy quarks, $Q=c,b,t$, their lowest-lying resonances can be described with the structure $Q\bar{q}\in \bm{\bar{3}}_f$.
The lowest-lying baryons can also be classified by SU$_f$(3) symmetry.

As the number of hadrons continues to increase over time, many resonances beyond the lowest-lying states have been
accumulated in the PDG~\cite{PDG24}.
The PDG includes a vast array of high-mass resonances,
which can lead to multiple possible descriptions.
First, these resonances can be described as excited states of the lowest-lying resonances.
In particular, high-mass mesons can be understood as orbital excitations of $q\bar{q}$
with angular momentum $\ell > 0$ or as radial excitations,
and are still expected to form flavor nonets with quantum numbers naturally corresponding to this description.

A different description, which is completely uncorrelated from the lowest-lying resonances,
is multiquarks---hadrons composed of four or more constituent quarks.
In the light-quark sector, there are long-standing candidates for tetraquarks, specifically the spin-0 nonet,
which we refer to as the $0^+_A$ nonet below,
composed of $a_0 (980)$, $K_0^* (700)$, $f_0 (500)$, and $f_0 (980)$~\cite{Jaffe77a, Jaffe77b, Jaffe04}. Additionally, around 500 MeV
above the $0^+_A$ nonet, there is another nonet,
referred to as the $0^+_B$ nonet,
composed of $a_0 (1450)$, $K_0^* (1430)$, $f_0 (1370)$, and $f_0 (1500)$,
which can also be considered as tetraquarks~\cite{Maiani:2006rq,Kim:2016dfq,Kim:2017yvd,Kim:2018zob,Lee:2019bwi}.
The $d^*(2380)$ resonance reported in Ref.~\cite{WASA-at-COSY:2011bjg} may be a hexaquark state~\cite{Kim:2020rwn}.
Multiquark candidates in the heavy-quark sector have also been reported in the literature,
including $\chi_{c1} (3872)$, $X^{\pm} (4020)$, $\chi_{c1} (4140)$,
$Z_c (3900)$~\cite{Belle03, BESIII:2013ouc, LHCb:2016axx, Xiao:2013iha}, $T_{cc}^+ (3875)$~\cite{LHCb:2021auc,LHCb:2021vvq},
$P_c (4312)$, $P_c (4440)$ and $P_c(4457)$~\cite{LHCb:2015yax,LHCb:2019kea}.

Alternatively, these multiquark candidates can also be described as hadronic molecules~\cite{Guo:2017jvc}.
For instance, the isoscalar resonance, $f_0(500)$, may be a meson molecule composed of $\pi\pi$~\cite{Ahmed:2020kmp}, while
the $f_0(1370)$ can be a $\rho\rho$ molecule~\cite{Molina:2008jw}.
The $a_0 (980)$ and $f_0 (980)$ might be interpreted as molecular states like $K\bar{K}$ or dynamically generated from $\pi\eta$ or
$K\bar{K}$~\cite{Weinstein:1990gu, Branz:2007xp, Branz:2008ha, Janssen:1994wn}.
The $\chi_{c1}$(3872) could be a meson molecular state composed
of $D\bar{D}^*$~\cite{Tornqvist:2004qy, Tornqvist:1993ng}, and
the $P_c (4312)$, $P_c (4440)$, $P_c$(4457) could be hadronic molecules, specifically
$\Sigma_c \bar{D}$ ($J^P=1/2^-$), $\Sigma_c \bar{D}^*$ ($J^P=3/2^-$), $\Sigma_c \bar{D}^*$ ($J^P=1/2^-$), respectively~\cite{Wu:2010jy, Wu:2010vk, Du:2019pij,Xiao:2019mvs}.
However, for the two nonets in the light quark sector mentioned above, it is not clear whether the molecular model can apply to all members~\cite{Kim:2023tph}.
Nevertheless, the structure of high-mass resonances remains open to various descriptions, making it necessary to pursue further efforts to
better understand them.

Among various resonances, those with $J^P=0^+$ in the light-quark sector are particularly
intriguing.
Currently, there appear to be three nonets in the PDG.
With the recent inclusion of the resonance $a_0(1710)$~\cite{BESIII:2021anf} in the PDG,
a third nonet in the $J^P=0^+$ channel, which we refer to as the $0^{+}_C$ nonet, can be identified as $a_0(1710)$, $K_0^*(1950)$, $f_0(1710)$, and $f_0(1770)$,
in addition to the $0^{+}_A$ and $0^{+}_B$ nonets mentioned above~\footnote{In our previous studies~\cite{Kim:2016dfq,Kim:2017yvd,Kim:2018zob,Lee:2019bwi},
the $0^+_A$ and $0^+_B$ nonets were referred to as the light and heavy nonets, respectively.}.
At first glance, the two-quark description seems appropriate, as the quantum numbers of these resonances
can be reproduced if they are viewed as orbital excitations of $q\bar{q}$ with an angular momentum of $\ell = 1$, i.e., $q\bar{q} (\ell = 1$) states.
The emergence of the three nonets in this description may be understood if they are distinguished by the radial excitations.

However, the main issue with this description is that the mass ordering expected from $q\bar{q} (\ell = 1)$ states
is opposite to the experimentally observed ordering, particularly for the $0^{+}_A$ nonet: $a_0 (980)$, $K_0^* (700)$, $f_0 (500)$, and $f_0 (980)$.
Indeed, the opposite mass ordering can be explained~\cite{Jaffe77a, Jaffe77b, Jaffe04} if the nonet members
are viewed as tetraquarks composed of diquark-antidiquark pairs.
The $0^{+}_B$ nonet, composed of
$a_0 (1450)$, $K_0^* (1430)$, $f_0 (1370)$ and $f_0 (1500)$, exhibits a similar ordering, and its members
can be considered as tetraquarks as well.
Eventually, through the tetraquark mixing model~\cite{Kim:2016dfq,Kim:2017yvd,Kim:2018zob,Kim:2023tph,Kim:2024adb},
the two nonets can be represented as linear combinations of two tetraquark types,
demonstrating that the two nonets can be treated as tetraquarks with different color-spin configurations.
From this consideration, the $0^{+}_C$ nonet may be treated as two-quark states, $q\bar{q} (\ell = 1)$.

Still, a remaining challenge is the marginal mass ordering observed in the $0^{+}_B$ nonet,
with $M[a_{0}(1450)] \gtrsim M[K_{0}^{*}(1430)]$, which is difficult to fully understand.
The marginal mass ordering contrasts with the clear mass ordering $M[a_{0}(980)] > M[K_{0}^{*}(700)]$ in the $0^{+}_A$ nonet.
Certainly, a mechanism is anticipated to generate the marginal mass ordering in the $0^{+}_B$ nonet without affecting the $0^{+}_A$ nonet.
Specifically, we need a mechanism that introduces an additional $q\bar{q} (\ell =1)$ component, as
it can reduce the mass gap generated by the tetraquark component.

A viable approach for this purpose is to develop a mixing mechanism between tetraquarks and $q\bar{q} (\ell = 1)$ states.
Indeed, Black {\it et al.} in Ref.~\cite{Black:1999yz} proposed a mixing between tetraquarks and $q{\bar q}(\ell=1)$ states,
which could result in the formation of two nonets: the $0^{+}_A$ nonet [$a_0 (980)$, $K_0^* (700)$, $f_0 (500)$, $f_0 (980)$],
and the $0^{+}_B$ nonet [$a_0 (1450)$, $K_0^* (1430)$, $f_0 (1370)$, $f_0 (1500)$].
While this mechanism can resolve the issue of marginal mass ordering in the $0^{+}_B$ nonet,
it also complicates the structure of the $0^{+}_A$ nonet, leading to $a_0(980)$ and $K_0^*(700)$ no longer being regarded as tetraquarks.
As will be discussed in Sec.~\ref{sec:Nonets}, this undermines the original motivation for introducing tetraquarks
to explain the mass ordering of the $0^{+}_A$ nonet.

Technically, the mixing mechanism proposed in Ref.~\cite{Black:1999yz} is formulated solely based on the flavor structure
of the tetraquarks being a nonet.
However, the color-spin structures of tetraquarks may also need to be taken into account in this consideration.
According to the tetraquark mixing model~\cite{Kim:2016dfq,Kim:2017yvd,Kim:2018zob,Kim:2023tph,Kim:2024adb},
the $0^{+}_A$ nonet has a different color-spin structure from the $0^{+}_B$ nonet, and this distinction must be accounted
for developing mixing mechanisms.
In principle, the three nonets mentioned above can be viewed as products of distinct mixing mechanisms.
Besides the mixing mechanism that produces the $0^{+}_A$ and $0^{+}_B$ nonets,
alternative mechanisms could also be considered, which might generate the $0^{+}_A$ and $0^{+}_C$ nonets or the $0^{+}_B$ and $0^{+}_C$ nonets.
The differences in color-spin structure may provide insights into constructing
a suitable mixing mechanism that addresses the marginal mass ordering in the $0^{+}_B$ nonet
while preserving the tetraquark nature of the $0^{+}_A$ nonet.
With this objective, we explore three possible mixing mechanisms and discuss the advantages and limitations associated with each.

This paper is organized as follows. Sec.~\ref{spectroscopy} reviews the resonances with $J^P=0^+$ listed in the PDG,
which are examined from two perspectives: two-quark states in Sec.~\ref{sec:two} and tetraquark states in Sec.~\ref{sec:tetraquark}.
We address the mass ordering issue in both descriptions and, eventually, motivates a mixing mechanism.
Sec.~\ref{sec:mechanism} reexamines the mixing mechanism
of Ref.~\cite{Black:1999yz} and discusses three mixing scenarios.
In Sec.~\ref{sec:result}, we provide the results
of the mixing mechanism that generates the $0^{+}_B$ and $0^{+}_C$ nonets.
We finally summarize in Sec.~\ref{sec:summary}.

\section{Spectroscopy of Mesons with $J^{P}=0^{+}$ }
\label{spectroscopy}

In the PDG~\cite{PDG24}, there are many resonances in the light quark system with the quantum numbers $J^{P}=0^{+}$, as shown
in Table~\ref{spin0+}. This table also lists the isospins, masses, and decay widths of the resonances.
These high-mass resonances in this table are subject to occasional updates, so their status can change over time.
For example, compared to the 2016 PDG~\cite{PDG16}, $a_0(1710)$ and $f_0(1770)$ have newly appeared.
The masses and decay widths evolve significantly with each new edition of the PDG.
In light of this, our analysis based on the resonances in this table may shift with future data updates.
Given this fluid situation, the structure of higher-mass resonances remains much less understood compared to that of the lowest-lying resonances
($J^P=0^{-}: \pi, K, \eta, \eta^\prime$; $J^P=1^{-}: \rho, K^*, \omega, \phi$).
In this section, we examine the resonances with $J^{P}=0^{+}$ from two perspectives:
the two-quark and tetraquark descriptions, to motivate a mixing mechanism.

\begin{table}
\begin{tabular}{c|c|c|c|c}  \hline
 $J^{PC}$ & $I$ &  Meson  & Mass(MeV) & $\Gamma$(MeV) \\
\hline
\multirow{14}{*}{$0^{++}$}
         &  0  & $f_0 (500)$  & 400$-$800 &  100$-$800   \\
         &  0  & $f_0 (980)$  & 990 &  10$-$100  \\
         &  1  & $a_0 (980)$  & 980 &  50$-$100   \\
         & 0 & $f_0 (1370)$ & 1200$-$1500 &  200$-$500   \\
         & 1 & $a_0 (1450)$ & 1439 &  258  \\
         & 0 & $f_0 (1500)$ & 1522 &  108   \\
         & 0 & $f_0 (1710)$ & 1733 &  150   \\
         & 1 & $a_0 (1710)$ & 1713 &  107   \\
         & 0 & $f_0 (1770)$ & 1784 &  161   \\
         & 1 & $a_0 (1950)$ & 1931 &  271 \\
         & 0 & $f_0 (2020)$ & 1982      & 436   \\
         & 0 & $f_0 (2100)$ & 2095 &  287   \\
         & 0 & $f_0 (2200)$ & 2187 &  210   \\
         & 0 & $f_0 (2330)$ & 2314(?) &  ?   \\
\hline
\multirow{3}{*}{$0^{+}$}
        & 1/2 & $K_0^* (700)$ & 845 & 468  \\
        & 1/2 & $K_0^* (1430)$ & 1425 & 270  \\
        & 1/2 & $K_0^* (1950)$ & 1957 & 170  \\
\hline
\end{tabular}
\caption{Resonances with $J^{P}=0^{+}$ are collected from the PDG~\cite{PDG24}.
Question marks (``?'') show that the listed values are either not clearly determined or not available from the current PDG.}
\label{spin0+}
\end{table}

\subsection{Two-quark description}
\label{sec:two}

One immediate observation is that
the isospins of all the resonances in Table~\ref{spin0+} are constrained to $I=0,1/2,1$, with no states having $I>1$.
This isospin composition is the same as that of the lowest-lying mesons, the pseudoscalar
($0^-$) nonet and the vector ($1^-$) nonet.
This suggests that most of these resonances may be grouped into flavor nonets, similar to the lowest-lying mesons.
This is, in fact, consistent with a view that these resonances with $J^P=0^+$ are described by
orbitally excited states of the lowest-lying mesons with $\ell=1$, namely, the $q\bar{q}(\ell=1)$ states with total angular momentum $J=0$,
\begin{eqnarray}
|0^{+}\rangle_2 = [(q\bar{q})_{S=1}\otimes (\ell=1)]_{J=0}\label{0+t}\ .
\end{eqnarray}
The subscript ``2'' on the left-hand side denotes two-quark states, included to distinguish them from tetraquark states with similar quantum numbers later.
By focusing on the isovector ($I=1$) and isodoublet ($I=1/2$) resonances in each channel,
and combining them with isoscalar resonances, we roughly assign from the resonances in the table three nonets as follows:
\begin{eqnarray}
0^{+}_A:&&~ a_0 (980), K_0^* (700), f_0 (500), f_0 (980)\label{0A}\  , \\
0^{+}_B:&&~ a_0 (1450), K_0^* (1430), f_0 (1370), f_0 (1500)\label{0B}\  , \\
0^{+}_C:&&~ a_0 (1710), K_0^* (1950), f_0 (1710), f_0 (1770)\label{0C}\  .
\end{eqnarray}
To further illustrate this, we note that, in Table~\ref{spin0+},
there are four members with $I=1$, three members with $I=1/2$, and many isoscalar resonances ($I=0$).
This leads to assign at least three nonets in the $0^{+}$ channel, which we distinguish by the subscripts $A$, $B$, and $C$.
There are enough isoscalar resonances to accommodate this assignment.
While our selection of isoscalar resonances is based on mass ordering along with the isovector members,
this is just one possibility, and other assignments could be considered for each nonet.
It should also be noted that some resonances in the table are still not included in these nonets.

One well-known issue is the mass ordering among the isospin members within each nonet introduced above.
For this discussion, we focus only on the mass ordering among the isovector ($I=1$) and isodoublet ($I=1/2$) members,
as the isoscalar members involve additional ambiguities, such as ideal mixing and anomalies.
Furthermore, $f_0(1370)$ and $f_0(1710)$ could be described alternatively as dynamically generated states~\cite{Geng:2008gx}.
If the $q\bar{q}(\ell=l)$ description holds,
one would anticipate the mass ordering in each nonet:
\begin{equation}
M[(q\bar{q})_{I=1}] < M[(q\bar{q})_{I=1/2}]\label{2qor}\ ,
\end{equation}
since the isovector member, consisting of a $u\bar{d}$ pair,
should be lighter than the isodoublet member, consisting of a $u\bar{s}$ pair.

However, as is well known, this mass ordering is not maintained in the $0^{+}_A$ nonet, as the isospin members exhibit
the opposite mass ordering:
\begin{equation}
M[a_0 (980)] > M[K_0^* (700)]\label{0Aor}\ .
\end{equation}
In the $0^{+}_B$ nonet, the isospin members show a marginal ordering:
\begin{eqnarray}
M[a_0(1450)] \gtrsim M[K^*_0(1430)]\label{0Bor}\ ,
\end{eqnarray}
which is also inconsistent with the expected ordering from the $q\bar{q}(\ell=l)$ structure.
Only the $0^{+}_C$ nonet members exhibit the expected ordering:
\begin{eqnarray}
M[a_0 (1710)] < M[K_0^* (1950)]\label{0Cor}\ .
\end{eqnarray}

\subsection{Tetraquark description}
\label{sec:tetraquark}

As is well known, to explain the mass ordering in Eq.~(\ref{0Aor}), the tetraquark structure is anticipated for the $0^+_A$ nonet
[$a_0 (980), K_0^* (700), f_0 (500), f_0 (980)$].
As suggested by Jaffe~\cite{Jaffe77a, Jaffe77b, Jaffe04}, the mass ordering in Eq.~(\ref{0Aor}) can be explained if
the $0^{+}_A$ nonet forms a tetraquark nonet constructed by combining spin-0 diquarks in ($\bm{\bar{3}}_c$, $\bm{\bar{3}}_f$) with their antidiquarks.
In this configuration, the isovector member ($\sim us\bar{s}\bar{d}$) is expected to be heavier than the isodoublet
member ($\sim ud\bar{d}\bar{s}$):
\begin{equation}
M[(qq\bar{q}\bar{q})_{I=1}] > M[(qq\bar{q}\bar{q})_{I=1/2}]\label{4qor}\ .
\end{equation}
Notice that this expectation is based on the assumption that the mass ordering is driven by the constituent quark mass, $m_q$,
with $m_s > m_u \approx m_d$.

The $0^{+}_B$ nonet [$a_0 (1450), K_0^* (1430), f_0 (1370), f_0 (1500)$] also exhibits a similar mass ordering, $M[a_0(1450)] \gtrsim M[K^*_0(1430)]$,
although the mass difference is marginal, $\Delta M \approx14$ MeV.
This ordering, nevertheless, motivates proposing a second type of tetraquark~\cite{Kim:2016dfq,Kim:2017yvd,Kim:2018zob},
formed by combining spin-1 diquarks in ($\bm{6}_c$, $\bm{\bar{3}}_f$) with their antidiquarks.
Including this, there are two types of tetraquarks in the $J^{P}=0^{+}$ channel, whose spin, color, and flavor structures are given by:
\begin{eqnarray}
&&~~~~~~~~~~~~~~~~\underline{\text{spin}}~~~~~\underline{\text{color}}~~~~~~~~~\underline{\text{flavor}}\nonumber \\
&&|\text{Type1} \rangle_4 = | 000 \rangle \otimes |\bm{1}_c \bar{\bm{3}}_c \bm{3}_c\rangle \otimes |\bm{9}_f \bar{\bm{3}}_f \bm{3}_f\rangle \label{type1}\ ,\\
&&|\text{Type2} \rangle_4 = | 011 \rangle \otimes |\bm{1}_c \bm{6}_c \bar{\bm{6}}_c\rangle \otimes |\bm{9}_f \bar{\bm{3}}_f \bm{3}_f\rangle \label{type2}\ ,
\end{eqnarray}
where the subscript ``4'' in the left-hand side denotes the tetraquarks.
Here, the first number represents the state of the tetraquark, the second number represents the state of the diquark, and the third number represents the state of the antidiquark.
For example, $|\bm{1}_c \bar{\bm{3}}_c \bm{3}_c\rangle$,
represents a color-singlet ($\bm{1}_c$) tetraquark composed of $\bar{\bm{3}}_c$ diquarks and $\bm{3}_c$ antidiquarks.
Note that these two types, which differ in color-spin structure but share the same flavor structure as a nonet, are orthogonal.
For a detailed explanation of these formulas, refer to Ref.~\cite{Kim:2024adb}.

These two tetraquark types, Eqs.(\ref{type1}) and (\ref{type2}),
have isospins constrained to $I=0, 1/2, 1$ similarly as the two-quark description above.
Note that tetraquarks with higher isospins ($I > 1$) could mathematically be constructed
by using other diquarks with the spin, color, flavor structure of ($J=1, \bar{\bm{3}}_c, \bm{6}_f$) or ($J=0, \bm{6}_c, \bm{6}_f$).
However, these diquarks are unstable due to repulsive binding from the color-magnetic force,
making the formation of tetraquarks with such diquark constituents highly unlikely.
This view is indeed consistent with
the fact that the resonances in Table~\ref{spin0+} is restricted to $I=0, 1/2, 1$
and the two tetraquark types can be used to describe the resonances in Table~\ref{spin0+}.

One important aspect, which was discussed in Refs.~\cite{Kim:2016dfq,Kim:2017yvd,Kim:2018zob}, is that
the two tetraquark types, Eqs.~(\ref{type1}) and (\ref{type2}),
are not eigenstates of the color-spin interaction, as they mix strongly through the color-spin interaction.
Instead, their linear combinations, which diagonalize the color-spin interaction, can be identified as
the two nonets, $0^{+}_A$ and $0^{+}_B$:
\begin{eqnarray}
|0^{+}_A \rangle_4 &=&~~\beta | \text{Type1} \rangle_4 + \alpha |\text{Type2} \rangle_4 \label{light}\ ,\\
|0^{+}_B \rangle_4 &=& -\alpha | \text{Type1} \rangle_4 + \beta |\text{Type2} \rangle_4 \label{heavy}\ .
\end{eqnarray}
The diagonalization also fixes the mixing parameters, $\alpha\approx \sqrt{{2}/{3}}$, $\beta\approx 1/\sqrt{3}$.
From these expressions, we can also verify that $|0^{+}_A \rangle_4$ and $|0^{+}_B \rangle_4$ are orthogonal.
This is the tetraquark mixing model~\cite{Kim:2016dfq,Kim:2017yvd,Kim:2018zob}.
This mixing model has several successful aspects, including the mass splitting between the $0^{+}_A$ and $0^{+}_B$,
their coupling strengths when decaying into two pseudoscalar mesons~\cite{Kim:2017yur,Kim:2022qfj,Kim:2023bac},
and hidden-color contributions~\cite{Kim:2024adb,Brink:1994ic}.

However, the marginal mass ordering in the $0^{+}_B$ nonet is still an issue within the tetraquark description.
Since $|0^{+}_A \rangle_4$ and $|0^{+}_B \rangle_4$, share the same flavor structure, a nonet,
both tetraquarks should exhibit the similar mass ordering as described in Eq.~(\ref{4qor}).
The actual situation is that, while this ordering is clearly satisfied in the $0^{+}_A$ nonet,
it is only marginally satisfied in the $0^{+}_B$ nonet.

Consequently, a mechanism is needed to generate marginal mass ordering in the $0^{+}_B$ nonet
without affecting the tetraquark structure of the $0^{+}_A$ nonet.
To develop this, we recall that the mass ordering in the $q\bar{q} (\ell = 1)$ nonet, as described by Eq.~(\ref{2qor}),
is opposite to the tetraquark ordering of Eq.~(\ref{4qor}).
The marginal mass ordering can be explained if the states involved contain the $q\bar{q} (\ell = 1)$
components in addition to the tetraquark components in their wavefunctions.
The opposite mass ordering from the $q\bar{q} (\ell = 1)$ component can reduce the mass gap between the isospin members,
resulting in the marginal ordering observed in the $0^{+}_B$ nonet.
A notable approach to producing two-component states is to utilize the mixing mechanism
between tetraquarks and $q\bar{q} (\ell = 1)$ states as in Ref.~\cite{Black:1999yz}.
Specifically, applying such a mixing mechanism between Eq.(\ref{0+t}) and Eq.(\ref{heavy}) can transform
\begin{eqnarray}
|0^{+}_B \rangle_4 &\Rightarrow& |0^{+}_B \rangle = a|0^{+}_B \rangle_4 + b|0^{+} \rangle_2 \label{heavy2}\ ,
\end{eqnarray}
where $a$ and $b$ are constants. This transformation can reduce the mass gap within the $0^{+}_B$ nonet.
The other nonet, $-b|0^{+}_B \rangle_4 + a|0^{+} \rangle_2$, also generated from this mechanism,
cannot be identified as the $0^+_A$ nonet since its tetraquark component is not consistent with the tetraquark structure of Eq.~(\ref{light}).
It is important to stress that the color-spin interaction remains diagonal in the basis of $|0^{+}_A \rangle_4$ [Eq.~(\ref{light})] and the new $|0^{+}_B \rangle$.

\section{Mixing mechanism }
\label{sec:mechanism}

In this section, we examine the mixing mechanism proposed by Black {\it et al.}~\cite{Black:1999yz} and explore how it can be
used to resolve the marginal mass ordering in the $0^{+}_B$ nonet.  To make our presentation explicit, we restate the introduction
by Black {\it et al.} for the mixing Lagrangian
between tetraquarks and the two-quark states of $q\bar{q}(\ell=1)$ and outline the technical steps for the mixing mechanism.
We then discuss the appropriate nonets where the mixing mechanism should be applied.

\subsection{Mixing Lagrangian}
\label{sec:mixing}

To develop a mixing mechanism, Black {\it et al.}~\cite{Black:1999yz} first represent the $q\bar{q}(\ell=1)$ nonet and
the tetraquark nonet by the tensors
\begin{eqnarray}
&&N^{\prime b}_a = q_a \bar{q}^b, ~~(q_a=u,d,s)\label{2nonet}\ ,\\
&&N^b_a = T_a \bar{T}^b\label{4nonet}\ ,
\end{eqnarray}
where diquarks ($T^a$) and antidiquarks ($\bar{T}^b$) are given as
\begin{eqnarray}
T_a &=&\frac{1}{\sqrt{2}}\epsilon_{abc}q^b q^c\equiv [q^b q^c]\ ,\nonumber\\
\bar{T}^a &=& \frac{1}{\sqrt{2}}\epsilon^{abc}\bar{q}_b \bar{q}_c \equiv [{\bar q}_b {\bar q}_c] \ .
\end{eqnarray}
If the nonet members, $N^{\prime 3}_3= s\bar{s}$, $N^3_3 = T_3 \bar{T}^3=[ud][\bar{u}\bar{d}]$, are identified as physical states,
Eqs.~(\ref{2nonet}) and (\ref{4nonet}) represent the nonets in ideal mixing. In this case, the isoscalar members
cannot be connected via SU$_f$(3) rotations from other members, implying that SU$_f$(3) symmetry is broken.

To introduce a mixing between these two tensors, Black {\it et al.} proposed a mixing Lagrangian,
\begin{eqnarray}
{\cal L}=-\gamma\text{Tr}(NN^\prime) \label{mlag1}\ ,
\end{eqnarray}
which constitutes the simplest invariant term constructed from the two tensors, $N^{b}_a$ and $N^{\prime b}_a$.
Note that this mixing Lagrangian is constructed solely based on the flavor structure being nonets,
independent of the color and spin structures of $N^{b}_a$ and $N^{\prime b}_a$.
Hence, the present mixing mechanism does not account for differences in the color-spin structures.
Another thing to mention is that, even though the mixing Lagrangian involving isoscalar resonances breaks SU$_f$(3) symmetry,
those involving the isovector ($N^{2}_1$ and $N^{\prime 2}_1$) and isodoublet ($N^{3}_1$ and $N^{\prime 3}_1$) resonances
retain an SU$_f$(3)-symmetric form.
Specifically, these terms remain consistent, whether derived from the SU$_f$(3) symmetric expression\footnote{Here, $(\bm{8}_f)^{b}_a$ and $(\bm{8}_f^\prime)^{b}_a$
represent the octet parts of $N^{b}_a$ and $N^{\prime b}_a$, respectively.},
$-\gamma\text{Tr}(\bm{8}_f\bm{8}_f^\prime)$, or from Eq.~(\ref{mlag1}).
Thus, the mixing mechanism for the isovector and isodoublet members introduces less ambiguity in the model.
Lastly, to maintain SU$_f$(3) symmetry, the isovector ($I=1$) and isodoublet ($I=1/2$) terms must be represented by the same mixing parameter, $\gamma$.

With the mixing Lagrangian, Eq.~(\ref{mlag1}),
Black {\it et al.} formulated a $2\times 2$ matrix for the mass matrix in each isospin channel.
Diagonalizing the mass matrix results in the physical states expressed by the two components,
tetraquark and $q\bar{q} (\ell=1)$ components.
Black {\it et al.} applied this mixing mechanism to the isovectors and isodoublets,
and generated the physical states,  $K_0^* (700), a_0 (980)$ in the $0^{+}_A$ nonet and $K_0^* (1430), a_0 (1450)$
in the $0^{+}_B$ nonet.

\subsection{Technical steps}
\label{sec:tech}

Here, we reformulate the technical steps of the mixing mechanism~\cite{Black:1999yz}.
To explain this comprehensively, we denote the states prior to mixing---the tetraquark and $q\bar{q}(\ell=1)$ states---as $|a\rangle$ and $|a^\prime\rangle$
and the states after mixing as $|A\rangle$ and $|A^\prime\rangle$, respectively.
Depending on the physical states corresponding to $|A\rangle$ and $|A^\prime\rangle$,
we can establish the corresponding mixing mechanism.

The mixing Lagrangian, Eq.~(\ref{mlag1}),
leads to a $2\times 2$ mass matrix, $M^2$, in the basis of $|a\rangle$ and $|a^\prime\rangle$, and this can be diagonalized as follow:
\begin{eqnarray}
M^2 =
\begin{array}{l|cc}
 & |a \rangle & |a^\prime \rangle \\
\hline
|a \rangle & m^2_{a} & \gamma \\
|a^\prime \rangle & \gamma & m^2_{a^\prime}
\end{array}
\quad
&\rightarrow&
\quad
\begin{array}{l|cc}
 & |A \rangle & |A^\prime \rangle \\
\hline
|A \rangle & M^2_{A} & 0 \\
|A^\prime \rangle & 0  & M^2_{A^\prime}
\label{mass matrix}\
\end{array}\ .
\end{eqnarray}
Here, $m_a$, $m_{a^\prime}$ are pre-mixing masses, and $M_{A}$, $M_{A^\prime}$ are physical masses after mixing.

\begin{figure}[t]
\centering
\epsfig{file=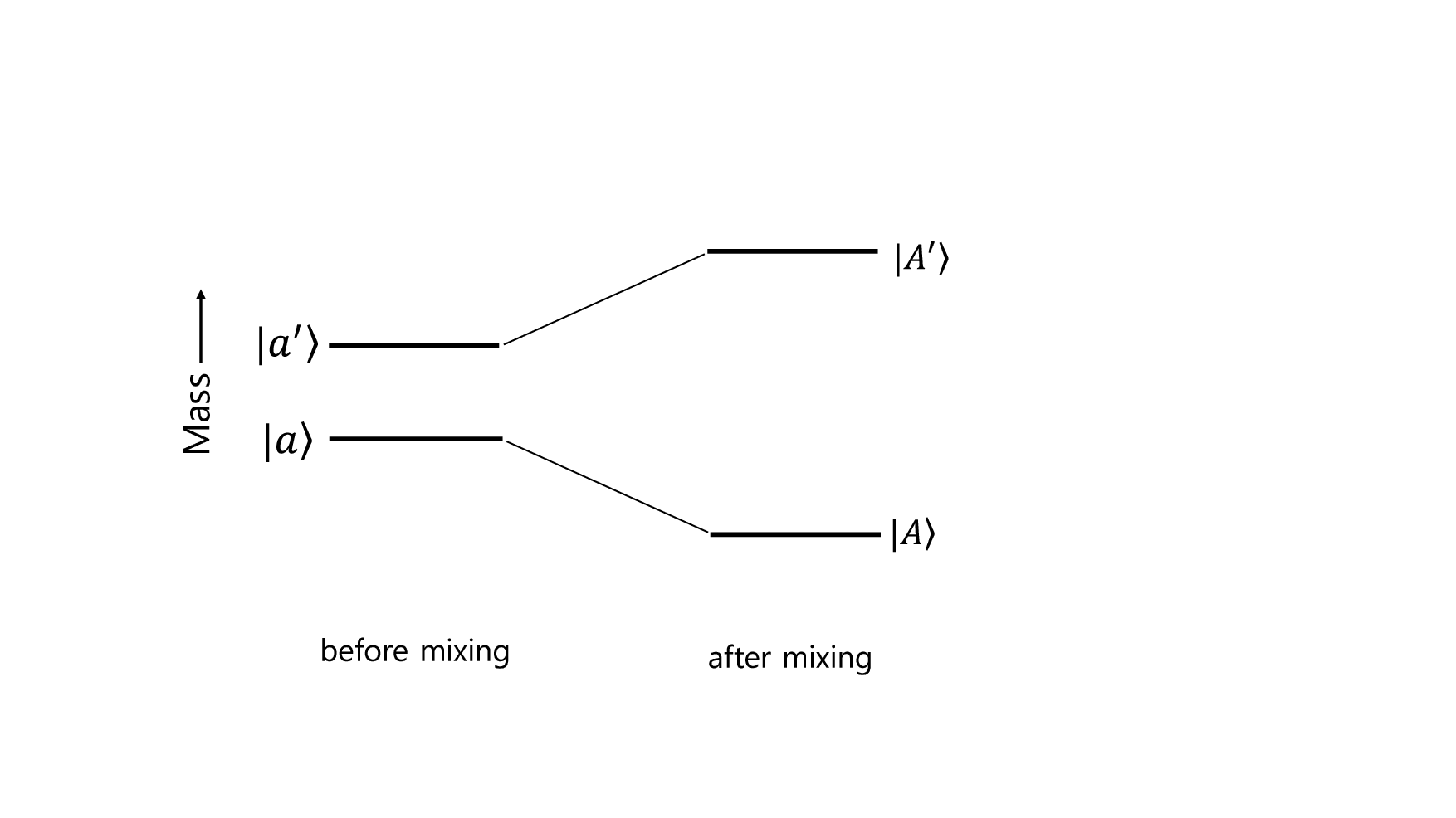, width=1.3\hsize}
\caption{A schematic diagram for the mixing mechanism. }
\label{mechanism}
\end{figure}

From the diagonalization process as in Eq.~(\ref{mass matrix}), the mass eigenvalues can be
expressed in terms of the unknowns, $m_a, m_{a^\prime}$ and $\gamma$ as follows:
\begin{eqnarray}
&&M^2_A= \frac{1}{2}\left [ m^2_{a^\prime}+m^2_a-\sqrt{(m^2_{a^\prime}-m^2_a)^2+4\gamma^2}\right ] \label{m1}\ ,\\
&&M^2_{A^\prime}= \frac{1}{2}\left [ m^2_{a^\prime}+m^2_a + \sqrt{(m^2_{a^\prime}-m^2_a)^2+4\gamma^2}\right ]\label{m2}\ .
\end{eqnarray}
Conversely, the pre-mixing masses can be determined from $M_A$ and $M_{A^\prime}$ as~\cite{Black:1999yz}:
\begin{eqnarray}
&&m^2_a= \frac{1}{2}\left [ M^2_{A^\prime}+M^2_A-\sqrt{(M^2_{A^\prime}-M^2_A)^2-4\gamma^2}\right ]\label{o1}\ , \\
&&m^2_{a^\prime}= \frac{1}{2}\left [ M^2_{A^\prime}+M^2_A + \sqrt{(M^2_{A^\prime}-M^2_A)^2-4\gamma^2}\right ]\label{o2}\ ,
\end{eqnarray}
which give the pre-mixing masses as functions of $\gamma$.
This mixing mechanism is illustrated schematically in Fig.~\ref{mechanism}.
The pre-mixing states ($|a \rangle$, $|a^\prime \rangle$) become ($|A \rangle$, $|A^\prime \rangle$) and,
consequently, the mixing increases the mass gap between the states.

The diagonalization also determines the physical states in terms of the pre-mixing states, $|a\rangle$ and $|a^\prime \rangle$ as:
\begin{eqnarray}
&&|A \rangle= C |a \rangle - D |a^\prime \rangle \label{A}\ ,\\
&&|A^\prime \rangle = D |a \rangle + C|a^\prime \rangle \label{AP}\ ,
\end{eqnarray}
where the corresponding coefficients are calculated to be:
\begin{eqnarray}
&&C = \frac{m^2_{a^\prime}- M^2_A}{\sqrt{(m^2_{a^\prime}-M^2_A)^2+\gamma^2}}\label{C}\ ,\\
&&D = \frac{M^2_{A^\prime}- m^2_{a^\prime}}{\sqrt{(m^2_{a^\prime}-M^2_{A^\prime})^2+\gamma^2}}\label{D}\  .
\end{eqnarray}
Thus, this mixing mechanism represents the physical states $|A \rangle$ and $|A^\prime \rangle$ in terms of tetraquarks and two-quark states of $q\bar{q}(\ell=1)$.

As a verification, we confirm that, when $\gamma=0$, Eqs.(\ref{o1}) and (\ref{o2})
lead to the expected results, $m^2_{a^\prime}=M^2_{A^\prime}$ and $m^2_{a}=M^2_{A}$, and
Eqs.(\ref{C}) and (\ref{D}) are simplified to $C=1$ and $D=0$.
As noted by Ref.~\cite{Black:1999yz}, the mixing parameter $\gamma$ has an upper bound under
the condition that $m^2_a$ and $m^2_{a^\prime}$, given in Eqs.~(\ref{o1}) and (\ref{o2}),
remain real values as long as the expression inside the square brackets is positive.

Specifically, the mixing parameter is maximized when
\begin{eqnarray}
 \gamma^2_{\text{max}}=\frac{(M^2_{A^\prime}-M^2_A)^2}{4}\label{mix max}\ ,
\end{eqnarray}
which leads to
\begin{eqnarray}
m^2_a = m^2_{a^\prime} = \frac{1}{2}[M^2_A + M^2_{A^\prime}]\label{cond1}\ .
\end{eqnarray}
From this, we note that
\begin{eqnarray}
&&m^2_{a^\prime}-M^2_A=\frac{1}{2}[M^2_{A^\prime}-M^2_A]=\gamma_{\text{max}} \nonumber ,\\
&&m^2_{a^\prime}-M^2_{A^\prime}=\frac{1}{2}[M^2_A - M^2_{A^\prime}]=-\gamma_{\text{max}}\nonumber\ ,
\end{eqnarray}
when $\gamma=\gamma_{\text{max}}$.
Substituting these expressions into Eqs.~(\ref{C}) and (\ref{D}), we obtain
\begin{eqnarray}
C=D=\frac{1}{\sqrt{2}}\label{CD}\label{cond2}\ .
\end{eqnarray}
As a result, when $\gamma=\gamma_{\text{max}}$, the physical states, $|A \rangle$ and $|A^\prime \rangle$, have
equal probabilities of being found in the tetraquark component and the $q\bar{q}(\ell=1)$ component.
In fact, the numerical factor, $1/\sqrt{2}$ in Eq.~(\ref{CD}), represents the maximum value of $D$, while it corresponds to
the minimum value of $C$. Hence,
\begin{eqnarray}
C\geq D\label{CD2}\label{cond3}\ ,
\end{eqnarray}
for any value of $\gamma$ within the range $0\leq \gamma \leq \gamma_{\text{max}}$.
Based on this, the physical state $|A\rangle$ has a greater tetraquark component
than the $q\bar{q}(\ell=1)$ component, while $|A^\prime\rangle$ has a greater $q\bar{q}(\ell=1)$ component.

In this mechanism, with fixed physical masses $M_A$ and $M_{A^\prime}$,
various sets of wave functions can be obtained depending on the value of $\gamma$.
Then the question is which value of $\gamma$ is most appropriate.
Ideally, the mixing parameter $\gamma$ should be independently determined by the underlying color-spin structures,
a topic beyond the scope of this work.
Instead, we examine the pre-mixing masses and resulting physical states, Eqs.~(\ref{A}) and (\ref{AP}),
as functions of $\gamma$ to identify the most appropriate value.

\subsection{Nonets Applicable for Mixing Mechanism}
\label{sec:Nonets}

The mixing Lagrangian in Eq.(\ref{mlag1}) depends solely on the flavor structures, $N^{b}_a$ and $N^{\prime b}_a$.
As is clear from the discussion in Sec.~\ref{sec:tech}, the mixing mechanism itself can be constructed independently of specific color-spin structures.
However, since quark pair annihilation is necessary for the mixing to occur, the mixing should,
in principle, depend on these structures~\cite{tHooft:2008rus}.
The issue lies in the fact that it remains unclear how to establish the mixing mechanism
based on the underlying color-spin configurations.

Without a concrete theoretical basis, the three nonets in the PDG---$0^{+}_A$, $0^{+}_B$, and $0^{+}_C$
[Eqs.~(\ref{0A}), (\ref{0B}), (\ref{0C})]---can be viewed as products of mixing mechanisms,
as each forms a flavor nonet.
Using the two-quark nonet of Eq.~(\ref{0+t}) and the tetraquark nonets in Sec.~\ref{sec:tetraquark},
one could consider a mixing mechanism that generates not only the $0^{+}_A$ and $0^{+}_B$ nonets,
as done by Ref.~\cite{Black:1999yz}, but also the $0^{+}_A$ and $0^{+}_C$ nonets, or the $0^{+}_B$ and $0^{+}_C$ nonets.
As the physical masses are used as inputs,
the relevancy of each mixing mechanism can be examined by analyzing the relationship
between the pre-mixing masses and the physical masses.
The mixing that generates the $0^{+}_B$ and $0^{+}_C$ nonets can be constructed by
combining $|0^{+}_B \rangle_4$ in Eq.~(\ref{heavy}) and $|0^{+}\rangle_2$ in Eq.~(\ref{0+t}).
This approach can resolve the issue of marginal mass ordering in the $0^{+}_B$ nonet
without affecting the tetraquark structure of the $0^{+}_A$ nonet.
It is also related to the mixing between $|0^{+}_A \rangle_4$ [Eq.~(\ref{light})] and $|0^{+}\rangle_2$,
which may lead to the formation of the $0^{+}_A$ and $0^{+}_C$ nonets.
Due to the differences in the color-spin structures, the tetraquark wave function $|0^{+}_A\rangle$ [Eq.(\ref{light})]
is orthogonal to $|0^{+}_B\rangle$ [Eq.(\ref{heavy})].
Thus, it is likely that the mixing generating the $0^{+}_A$ and $0^{+}_C$
nonets is minimized when the maximal mixing occurs to form the $0^{+}_B$ and $0^{+}_C$ nonets.
As can be seen in Sec.~\ref{sec:result}, maximal mixing indeed appears essential to occur for the $0^{+}_B$ and $0^{+}_C$ nonets.
Conversely, if maximal mixing occurs to generate the $0^{+}_A$ and $0^{+}_C$ nonets, resulting in minimal mixing for the $0^{+}_B$ and $0^{+}_C$ nonets, this does not
contribute to resolving the marginal mass ordering in the $0^{+}_B$ nonet.

The other mixing mechanism producing the $0^{+}_A$ and $0^{+}_B$ nonets,
which was proposed by Ref.~\cite{Black:1999yz}, could provide
an another solution to the marginal mass ordering in the $0^{+}_B$ nonet.
Such mixing can be constructed by combining the two-quark wave function $|0^{+}\rangle_2$ with a tetraquark wave function,
generating the $0^{+}_A$ nonet on the one hand and the $0^{+}_B$ nonet on the other.
Thus, this mixing involves only a single type of tetraquark.
As such, this is not compatible with the tetraquark mixing model~\cite{Kim:2016dfq,Kim:2017yvd,Kim:2018zob,Kim:2023tph,Kim:2024adb}
represented by Eqs.(\ref{light}) and (\ref{heavy}), which involves two distinct tetraquark structures.
Although it provides a rationale for the marginal mass ordering between $K_0^* (1430)$ and $a_0 (1450)$,
this mixing mechanism also alters the structure of $K_0^* (700)$ and $a_0 (980)$, diluting their tetraquark nature
by introducing a significant $q\bar{q} (\ell=1)$ component.
This undesirable aspect is evident from Eqs.(3.11) and (3.12) of Ref.~\cite{Black:1999yz},
\begin{eqnarray}
&&|a_0(980) \rangle ~= 0.707 |us\bar{s}\bar{d}\rangle - 0.707 |u\bar{d} \rangle\ ,\\
&&|K_0^* (700) \rangle = 0.857 |ud\bar{d}\bar{s}\rangle - 0.515 |u\bar{s} \rangle\ ,
\end{eqnarray}
obtained at the maximum mixing parameter, $\gamma^2_{\text{max}}=0.33$ GeV$^4$.
These show that the $a_0(980)$ state has equal proportions of both the tetraquark component and the $q\bar{q}(\ell=1)$ component.
The $K_0^* (700)$ state also acquires significant contribution from the $q\bar{q}(\ell=1)$ component.
As a result, the $a_0 (980)$ and $K_0^* (700)$ generated from the mixing cannot be regarded as tetraquarks anymore.

Furthermore, using these wave functions,
one can easily calculate the main quark-mass, $m_q$, contribution to the masses of $a_0(980)$ and $K_0^* (700)$.
It can be shown that the $m_q$ contribution to the $K_0^* (700)$ mass is larger than that to the $a_0(980)$ mass,
by approximately $[2(0.857)^2-1]m_u=0.47m_u > 0$ where $m_u$ is the constituent $u$-quark mass.
Thus, the original motivation for tetraquarks, which is based on the mass ordering $M(us\bar{s}\bar{d}) > M(ud\bar{d}\bar{s})$,
as conjectured from the $m_q$ contribution (see the discussion in Sec.~\ref{sec:tetraquark}), is no longer well supported.
In other words, the experimental mass ordering of $M[a_0 (980)] > M[K_0^* (700)]$ is largely attributed to the mixing mechanism,
which is independent of the fact that $m_s > m_u \approx m_d$.

Instead, in this work, we propose the mixing mechanism
generating the $0^{+}_B$ and $0^{+}_C$ nonets as a potential solution to the marginal mass ordering in the $0^{+}_B$ nonet.
With this mixing, we can maintain the widely accepted tetraquark interpretation of the $0^{+}_A$ nonet~\cite{Jaffe77a, Jaffe77b, Jaffe04, Maiani:2004uc},
and naturally explain the appearance of
the $0^{+}_C$ nonet in the PDG.
We summarize the isospin channels, pre-mixing configurations, and the physical states that result from this mixing:
\begin{eqnarray}
\begin{array}{l|l|l}
 I           & \text{pre-mixing}~      &~ \text{after mixing} \\
\hline
1           & |0^{+}_B\rangle_4, |0^{+}\rangle_2   &~  a_0 (1450), a_0 (1710) \\
\frac{1}{2} & |0^{+}_B\rangle_4, |0^{+}\rangle_2   &~  K_0^* (1430), K_0^* (1950)\
\end{array}
\label{mixing channels}
\end{eqnarray}
Here, we have two mixing channels: one for the isovector ($I=1$) members and the other for the isodoublet ($I=1/2$) members
but their mixing parameter $\gamma$ must be kept the same for both channels.
The configurations prior to mixing are given in Eqs.~(\ref{0+t}) and (\ref{heavy}).
Both form separate flavor nonets and can therefore be represented by tensors with flavor structures,
$N^b_a$ and $N^{\prime b}_a$ in Eqs.~(\ref{2nonet}) and (\ref{4nonet}).
The mixing Lagrangian is expressed as in Eq.(\ref{mlag1}), and the color-spin structure is not explicit in the calculation.
If this mixing mechanism is valid, the color-spin structure should be inherently reflected in the physical masses used as inputs.

\section{Mixing between $|0^{+} \rangle_2$ and $|0^{+}_B\rangle_4$}
\label{sec:result}

Using the technical procedure outlined in Sec.~\ref{sec:tech}, we now apply the mixing mechanism between
$|0^{+} \rangle_2$ and $|0^{+}_B\rangle_4$, and generate the physical states after mixing,
as indicated in Eq.~(\ref{mixing channels}).
For the $I=1$ channel, we identify $|A\rangle$ and $|A^\prime\rangle$ in Eqs.~(\ref{A}) and (\ref{AP})
as $a_0(1450)$ and $a_0(1710)$, respectively.  Denoting the corresponding coefficients as $C_a$ and $D_a$,
the mixing mechanism yields the wave functions for $a_0(1450)$ and $a_0(1710)$:
\begin{eqnarray}
&&|a_0(1450) \rangle= C_a |0^{+}_{B}\rangle_4 - D_a |0^{+}\rangle_2\ , (I=1)\label{0+I1}\ ,\\
&&|a_0(1710) \rangle = D_a |0^{+}_{B} \rangle_4 + C_a |0^{+}\rangle_2\ , (I=1)\label{0+I2}\ ,
\end{eqnarray}
where ``($I=1$)'' specifies that we are considering the isovector members from the nonets, $|0^{+}\rangle_2$ and $|0^{+}_{B}\rangle_4$,
in Eqs.~(\ref{0+t}) and (\ref{heavy}).
The physical masses, $M[a_0(1450)] = 1.439$ GeV and $M[a_0(1710)] = 1.713$ GeV are used as inputs
to determine the pre-mixing masses, ($m_a$, $m_{a^\prime}$),
as well as the coefficients $C_a$ and $D_a$, according to Eqs.~(\ref{o1}), (\ref{o2}), (\ref{C}), and (\ref{D}),
as functions of the mixing parameter $\gamma^2$.
In this case, the maximum mixing parameter calculated from Eq.~(\ref{mix max}) is
\begin{eqnarray}
\gamma^2_{\text{max}}=0.186~~ \text{GeV}^4\ ,
\end{eqnarray}
which is much smaller than $0.33$ GeV$^4$ of Ref.~\cite{Black:1999yz} obtained from other mixing scenario.
Our results are shown in the left part of Table~\ref{0+Re} under ``$I=1$'' for some chosen values of $\gamma^2$ within the range
$0\le \gamma^2 \le \gamma^2_{\text{max}}$.
The results from $\gamma^2_{\text{max}}=0.186$ GeV$^4$, which appear to be close to the realistic situation,
are presented in the last row of Table~\ref{0+Re}.

As $\gamma^2$ increases from 0 to $0.186$ GeV$^4$, we observe that
$C_a$ decreases from $1$ to $0.707$, while $D_a$ increases from 0 to $0.707$, consistent with the predictions of Eq.~(\ref{cond3}).
At the same time, $m_a$ increases from $1.439$ GeV to $1.582$ GeV, while $m_{a^\prime}$ decreases from $1.713$ GeV to the same
value of $1.582$ GeV, as expected from Eq.~(\ref{cond1}).
At small $\gamma^2$, the $|a_0(1450) \rangle$ wave function in Eq.~(\ref{0+I1}) is dominated by the tetraquark component $|0^{+}_{B}\rangle_4$,
while the $|a_0(1710) \rangle$ wave function is dominated by the two-quark component $|0^{+}\rangle_2$.
This dominance becomes reduced as $\gamma^2$ increases.
When $\gamma^2=\gamma^2_{\text{max}}$, the expected results, $C_a=D_a=0.707$, are achieved,
indicating that $|a_0(1450) \rangle$ and $|a_0(1710) \rangle$ have equal probabilities
of being found in both the tetraquark component and the two-quark component.

In fact,
Eqs.~(\ref{0+I1}) and (\ref{0+I2})
can provide an alternative constraint on the mixing parameter, $\gamma$.
When $a_0(1450)$ and $a_0(1710)$ decay into, for example, $\pi \eta$,
the associated coupling strengths can be calculated as $\langle \pi \eta |a_0(1450)\rangle$ and $\langle \pi \eta |a_0(1710)\rangle$, respectively.
Since the tetraquark component and the two-quark component have opposite relative signs in Eqs.(35) and (36),
the coupling strengths for $a_0(1450)$ and $a_0(1710)$ are expected to differ significantly, with one being small and the other large.
The difference in couplings is maximized when $\gamma = \gamma_{\text{max}}$, as the two components are equally probable.
This difference should manifest in the corresponding decay partial widths and could provide
a physical test to determine the $\gamma$ parameter.
However, to make this analysis feasible, precise experimental values for the partial decay widths are required, along with
a reliable framework for calculating $\langle \pi \eta |0^+_B\rangle_4$ and $\langle \pi \eta |0^+\rangle_2$.


\begin{table}
\centering
\begin{tabular}{c|c|c|c|c|c|c|c|c}  \hline
\multirow{2}{*}{$\gamma^2$ } &\multicolumn{4}{c|}{$I=1$}& \multicolumn{4}{c}{$I=1/2$} \\
\cline{2-9}
 & $C_a$ & $D_a$ & $m_{a}$ & $m_{a^\prime}$ & $C_K$ & $D_K$ & $m_{K}$ & $m_{K^\prime}$ \\
\hline
   $0$   & $1$ & $0$ & $1.439$ & $1.713$ & $1$ &  $0$ & $1.425$ & $1.957$  \\
 $0.05$  & $0.963$ & $0.269$ & $1.461$ & $1.695$ & $0.992$ & $0.125$ & $1.435$ & $1.950$   \\
 $0.10$   & $0.917$ & $0.399$ & $1.486$ & $1.672$ & $0.984$ & $0.179$ & $1.445$ & $1.942$  \\
 $0.15$  & $ 0.849 $ & $0.528$ & $1.520$ & $1.641$ & $0.975$ & $0.221$ & $1.455$ & $1.935$ \\
 $0.17$  & $ 0.805 $ & $0.593$ & $1.541$ & $1.622$ & $0.972$ & $0.236$ & $1.460$ & $1.931$ \\
 $0.186$  & $0.707$ & $0.707$ & $1.582$ & $1.582$ & $0.969$  &  $0.248$ & $1.463$ & $1.929$\\
\hline
\end{tabular}
\caption{Here, we present the coefficients of the mixing wave functions, with one set, $C_a$ and $D_a$,
corresponding to the $I=1$ case, and another set, $C_K$ and $D_K$, for the $I=1/2$ case.
Additionally, we provide the pre-mixing masses, denoted as $m_a$ and $m_{a^\prime}$ for the $I=1$ members,
and $m_K$ and $m_{K^\prime}$ for the $I=1/2$ members.
These values are calculated for several mixing parameters chosen within the range $0 \leq \gamma^2 \leq \gamma_{\text{max}}^2$.
The unit of $\gamma^2$ is GeV$^4$, while the units for $m_a$, $m_{a^\prime}$, $m_K$, and $m_{K^\prime}$ are GeV.
The last row shows the results for $\gamma^2 = \gamma_{\text{max}}^2=0.186$ GeV$^4$.
}
\label{0+Re}
\end{table}

For the $I=1/2$ channel, we identify $|A\rangle = K_0^*(1430)$ and $|A^\prime\rangle = K_0^*(1950)$,
and their corresponding wave functions are given as:
\begin{eqnarray}
&&|K_0^*(1430) \rangle= C_K |0^{+}_{B}\rangle_4 - D_K |0^{+}\rangle_2\ , (I=1/2)\label{0+I3}\ ,\\
&&|K_0^*(1950) \rangle = D_K |0^{+}_{B} \rangle_4 + C_K |0^{+}\rangle_2\ ,(I=1/2)\label{0+I4}\ .
\end{eqnarray}
Here, $C_K$ and $D_K$ differ from those in the isovector case of Eqs.~(\ref{0+I1}) and (\ref{0+I2}),
as the input masses are different.
The maximum mixing parameter in the $I=1/2$ channel, calculated from Eq.~(\ref{mix max}),
is $\gamma^2_{\text{max}}=0.809$ GeV$^4$, which is
significantly higher than that in the $I=1$ channel.  However, this is merely a nominal maximum and the mixing mechanism
for the $I=1/2$ channel should also be restricted within the same range as the $I=1$ channel,  $0 \leq \gamma^2 \leq 0.186$ GeV$^4$.
This is because both isovector and isodoublet terms in the mixing Lagrangian, Eq.~(\ref{mlag1}), should be described by
the same mixing parameter $\gamma$; otherwise, the isovector and isodoublet parts in the mixing Lagrangian would not be related by SU$_f$(3) symmetry.
Since the range of $\gamma^2$ is much less than the nominal maximum value, $0.809$ GeV$^4$, the $I=1/2$ results
are close to the no-mixing case ($\gamma=0$). This suggests that the state $|K_0^*(1430) \rangle$ in Eq.~(\ref{0+I3}) is dominated by
the tetraquark component $|0^{+}_{B}\rangle_4$, while $|K_0^*(1950) \rangle$ in Eq.~(\ref{0+I4}) is mainly described
by the two-quark component $|0^{+}\rangle_2$.
Even at the maximum, $\gamma^2=0.186$ GeV$^4$, as shown in the last row of the table, $C_K^2\approx 0.94$ and $D_K^2\approx 0.06$,
which is clearly in contrast with the isovector ($I=1$) case, $C_K^2=D_K^2=0.5$.
This certainly offers a different perspective on the structure of the $0^{+}_B$ and $0^{+}_C$ nonets
in Eqs.~(\ref{0B}) and (\ref{0C}).

In fact, the results for the large mixing parameters, close to the maximum $\gamma^2_{\text{max}}\approx 0.186$ GeV$^4$,
appear to reflect the physical expectations when the two results in the $I=1$ and $I=1/2$ channels are compared.
When $\gamma^2=0.186$ GeV$^4$, the pre-mixing masses are $m_a=1.582$ GeV and $m_K=1.463$ GeV, resulting in $m_a$ being greater than
$m_K$ by $\Delta m = 119$ MeV.
This positive mass difference qualitatively aligns with the expectation,
as the corresponding pre-mixing states are tetraquarks with the mass ordering given by Eq.~(\ref{4qor}).
Its magnitude, $\Delta m = 119$ MeV, is a reasonable value that can be understood from the quark mass difference, $m_s - m_u$.
After the mixing, the pre-mixing masses become the physical masses, $M[a_0(1450)]=1.439$ GeV and $M[K^*_0(1430)]=1.425$ GeV,
leading to the marginal gap, $\Delta M\approx 0.014$ GeV.
Therefore, this gap seen in the physical masses can be understood as a consequence of the mixing mechanism
that significantly reduces the isovector mass, $1.582\rightarrow 1.439$ GeV,
while only slightly decreasing the isodoublet mass, $1.463\rightarrow 1.425$ GeV.
The significant change in the isovector mass actually reflects a large contribution from the two-quark component
in the $|a_0(1450) \rangle$ wave function [Eq.~(\ref{0+I1})], generated through the mixing mechanism.
In contrast, the small change in the isodoublet mass
arises from the fact that the $|K_0^*(1430) \rangle$ wave function, in Eq.~(\ref{0+I3}), remains predominantly tetraquark in nature, even after mixing.

On the other hand, the results for small values of $\gamma^2$ do not align with the physical expectations.
For instance, when $\gamma^2=0.05$ GeV$^4$, the pre-mixing masses are $m_a=1.461$ GeV and $m_K=1.435$ GeV,
resulting in a mass gap of $\Delta m = 26$ MeV. This gap is too small to be adequately explained by the tetraquark mass ordering.
Even with a larger mixing, such as $\gamma^2=0.17$ GeV$^4$, the mass gap increases to $\Delta m = 81$ MeV, which is still somewhat insufficient.
Therefore, the results for $\gamma^2$ near its maximum appear more suitable.
Although this analysis does not specify the precise value of $\gamma^2$, the mixing mechanism at
$\gamma^2_{\text{max}}$ can accommodate
two nonets in the PDG, $0^{+}_B$ and $0^{+}_C$ [Eqs.~(\ref{0B}) and (\ref{0C})].

Another remark is that the resonance $a_0(1710)$, which is used in this analysis, was included in the PDG only recently.
Without this resonance, the mixing mechanism in the $I=1$ channel would have to rely on a much heavier resonance, $a_0(1950)$.
In that case, the maximum mixing parameter, calculated from Eq.~(\ref{mix max}), would become significantly larger than $0.186$ GeV$^4$,
making the conclusion above unattainable.

Finally, it is worthwhile to briefly mention the masses, $m_{a^\prime}$ and $m_{K^\prime}$, of pre-mixing two-quark states
at the maximum mixing parameter, $\gamma^2=0.186$ GeV$^4$.
As shown in Table~\ref{0+Re}, their masses satisfy the ordering $m_{K^\prime} > m_{a^\prime}$,
consistent with the two-quark description, despite the large mass gap, $\Delta m^\prime = m_{K^\prime} - m_{a^\prime} \approx 447$ MeV.
In fact, a better description might be obtained if a new $K_0^*$ resonance with a mass near 1.8 GeV were present.
Should this hypothetical resonance replace the $K_0^*(1950)$ in the $0^{+}_C$ nonet, our mixing mechanism would yield
a tetraquark mass gap of around $\Delta m \approx 100$ MeV and a two-quark gap of approximately $\Delta m^\prime \approx 170$ MeV.
Given the fluid situation of the PDG listings, this expectation for a new resonance in the future is not unrealistic.

In conclusion, a mixing mechanism has been developed to explain the marginal mass ordering seen in the $0^{+}_B$ nonet.
This mechanism, which involves the mixing of two-quark states and tetraquarks, generates the two nonets, $0^{+}_B$ and $0^{+}_C$, in the PDG.
Using the physical masses as inputs, this approach provides the pre-mixing masses and the wave functions after mixing.
For mixing parameters near maximal mixing, the pre-mixing tetraquark states are separated by a reasonable mass gap,
$\Delta m\approx 119$ MeV, between the isovector and isodoublet members.
Through the mixing, these states acquire two-quark components. The marginal mass ordering in the $0^{+}_B$ nonet appears
to result from the two-quark component, which develops strongly in the isovector channel but weakly in the isodoublet channel.

\section{Summary}
\label{sec:summary}

In this work, a mixing mechanism between two-quark states and tetraquarks is proposed as a
potential solution to the marginal mass ordering in the $0^{+}_B$ nonet, which comprises the
$a_0 (1450)$, $K_0^* (1430)$, $f_0 (1370)$ and $f_0 (1500)$.
Originally introduced in Ref.~\cite{Black:1999yz}, this approach enables the construction of any two nonets of interest
using two-quark and tetraquark components.
Although this mechanism depends solely on the flavor structure of tetraquarks,
the underlying color-spin structures should also be considered.
Indeed, there are three nonets, $0^{+}_A$, $0^{+}_B$, and $0^{+}_C$, as shown in Eqs.~(\ref{0A}), (\ref{0B}), (\ref{0C}),
that can be identified in the $J^P=0^{+}$ channel of the PDG.
According to the tetraquark mixing model, the $0^{+}_A$ and $0^{+}_B$ nonets have different color-spin structures, which
can give a clue for constructing an appropriate mixing mechanism among the three nonets.
We argued that the mixing mechanism that generates the $0^{+}_B$ and $0^{+}_C$ nonets is promising as it resolves
the marginal mass ordering in the $0^{+}_B$ nonet, maintains the widely accepted tetraquark interpretation of the
$0^{+}_A$ nonet, and explains the appearance of the $0^{+}_C$ nonet.
Other mixing mechanisms are found to be conflicting either with this mixing mechanism or with the tetraquark mixing model.

Indeed, using physical masses as inputs, we found that the mixing mechanism generating the $0^{+}_B$ and $0^{+}_C$ nonets
leads to a reasonable solution when the mixing parameter is near its maximum.
The pre-mixing masses of the isovector and isodoublet members are well separated, as expected from the tetraquark mass ordering.
We believe that this type of mixing can provide a new insight on the high-mass resonances in the PDG.
Future research may be needed to explore similar mixing in other spin channels or in the resonances of heavy quark sectors.

\acknowledgments
This work was supported by the National Research Foundation of Korea(NRF) grant funded by the
Korea government(MSIT) (No. NRF-2023R1A2C1002541, No. NRF-2018R1A5A1025563).

\end{document}